\documentclass[
reprint,
superscriptaddress,
amsmath,
amssymb,
aps,
prb,
floatfix,
]{revtex4-1}

\usepackage{graphicx}
\usepackage{dcolumn}
\usepackage{bm}
\usepackage[colorlinks,linkcolor=blue,citecolor=blue,urlcolor=blue]{hyperref}

\usepackage[raggedright,nooneline,large]{subfigure}  
\begin{document}
\preprint{APS/123-QED}
\title{Temperature dependent
determination of electron heat capacity and electron-phonon
factor for Fe$_{0.72}$Cr$_{0.18}$Ni$_{0.1}$}
\author{Jan Winter}
\email{jan.winter@hm.edu}
\affiliation{Munich University of Applied Sciences, Lothstrasse 34, 80335 Munich, Germany}
\author{J\"urgen Sotrop}
\affiliation{Munich University of Applied Sciences, Lothstrasse 34, 80335 Munich, Germany}
\author{Stephan Borek}
\affiliation{Ludwig Maximilian University Munich, Chemie Department, Butenandtstrasse 5-13, 81377 Munich, Germany}
\author{Heinz P. Huber}
\affiliation{Munich University of Applied Sciences, Lothstrasse 34, 80335 Munich, Germany}
\author{Jan Min\'ar}
\affiliation{Ludwig Maximilian University Munich, Chemie Department, Butenandtstrasse 5-13, 81377 Munich, Germany}
\affiliation{New Technologies-Research Center, University of West Bohemia, Univerzitni 8, 306 14 Pilsen, Czech Republic}
\date{Jun 2015}
\begin{abstract}
A theoretical approach using ab initio calculations has been applied to study the interaction of an ultra-short laser pulse with the metal alloy Fe$_{0.72}$Cr$_{0.18}$Ni$_{0.1}$ (AISI 304). The electronic structure is simulated by taking into account the chemical and magnetic disorder of the alloy by the coherent potential approximation implemented in a fully relativistic Korringa-Kohn-Rostoker-formalism in the framework of spin density functional theory. Utilizing these predictions we determined the electron heat capacity and the electron-phonon coupling factor of  Fe$_{0.72}$Cr$_{0.18}$Ni$_{0.1}$ in dependence on the electron temperature for two-temperature model applications. Compared with pure Fe a maximum deviation of 5~\% for the electron heat capacity and 25~\% for the electron-phonon coupling factor is found.
\end{abstract}
\maketitle
\section{Introduction}

Ultra-short laser pulses have a broad range of industrial applications and have been used for surface modifications of solid materials and precise micromachining\cite{Burakov,bauerleu,Hu}. In order to further optimize the laser process a deep understanding of the dynamics behind the laser ablation is essential. The theoretical description of laser-matter interaction demands a predictive modelling in irradiating a material target with an ultra-short laser pulse. Ultra-fast laser irradiation of a target material induces the electron and lattice temperature into a non-equilibrium state. For metals, the non-equilibrium processes induced in a target during the energy deposition can be described by the two-temperature model (TTM)\cite{S.I.AnisimovB.L.KapeliovichT.L.Perelman,S.I.AnisimovB.Rethfeld}. The TTM can be written as the energy balance of two coupled non-linear differential equations, which describe the spatial and temporal evolution of the electron and lattice temperatures: 
\begin{equation}
{C_e(T_e)\frac{\partial T_e}{\partial t} = \nabla \cdot [K_e(T_e,T_l)\nabla T_e] - G(T_e)(T_e-T_l)+S,}
\label{eq:TTM-1}
\end{equation}
\begin{equation}
{C_l(T_l)\frac{\partial T_l}{\partial t} = \nabla \cdot [K_l(T_l)\nabla T_l] + G(T_e)(T_e-T_l).} 
\label{eq:TTM-2}
\end{equation}
where $C$ is the heat capacity and $K$ is the thermal conductivity with respect to the temperature of the electron and lattice denoted by subscripts $e$ and $l$, $G$ is the electron-phonon coupling factor, and S is the laser heating source term.

The model is based on the absorption of the laser pulse energy by the valence band electrons and energy transfer from the hot electrons to the lattice vibrations due to electron-phonon coupling by an energy relaxation processes. The heat diffusion from this irradiated hot surface into the bulk material of the target is described by Fourier law. In the majority of cases, for ultra-short laser pulses the lattice heat conduction described in Eq.~(\ref{eq:TTM-2}) can be neglected in comparison to the electron heat conduction in metals. Based on this, the absorption of ultra-short laser pulses by the electrons result in a transient, strong non-equilibrium of electrons and lattice due to the small heat capacity of the lattice, and the long thermalization times of both electrons and lattice. The laser energy deposited in the irradiated metal target is stored in the electron subsystem while the lattice remains at a considerably low temperature. Accordingly, the electron temperature can increases by up to $\sim10^5$~K, such that there is a temperature dependency of the thermophysical parameters of the target material included in the TTM equations. Therefore, in order to obtain accurate results in the determination of these parameters, i.e. the electron heat capacity $C_e(T_e)$, electron-phonon coupling factor $G(T_e)$ and electron thermal conductivity $K_e(T_e)$, the temperature dependence cannot be neglected when using calculation methods.

In the beginning the first models for the description of laser ablation process were based on the free electron gas. More sophisticated models took the electronic structure from ab initio calculation\cite{Lin&Zhigilei&Celli}. The application of these models is limited to the description of metals. We extend these theoretical considerations to include chemical and magnetic disorder into the calculation of the electronic structure using the coherent potential approximation. This is essential for the description of technical relevant alloys like stainless steel.

It is well known that for transition metals the main electronic and magnetic properties can be ascribed to the $d$-band electrons. For the thermophysical behavior the $d$-band has an essential influence due to the thermal excitation of low-lying $d$-band electrons\cite{Raynor}. Therefore the $d$-band has to be taken into account for the calculation of the electron heat capacity and the electron-phonon coupling factor. The electron temperature dependence of thermophysical parameters and the affects by thermal excitation of $d$-band electrons have been analyzed by Lin~\textit{et~al.}\cite{Z.LinL.V.Zhigilei,Lin&Zhigilei}. This analysis connected the electronic structure calculations to the temperature dependence of the electron heat capacity and electron-phonon coupling factor. The results were compared with the free electron gas approximation and was related especially to the thermal excitation of $d$-band electrons. Previous work has focused on detailed calculations of the electron temperature dependency of thermophysical parameters, in particular of electron heat capacity and electron-phonon coupling factor of noble and other transition metals based on the analysis of the electronic structure by Lin~\textit{et~al.}\cite{Lin&Zhigilei&Celli}. Nevertheless, there is still a lack of knowledge for these basic thermophysical electronic data, especially for industrial relevant materials like metal alloys, e.g. stainless steel.

In this paper we present the electron temperature dependencies of electron heat capacity and electron-phonon coupling factor for the stainless steel alloy Fe$_{0.72}$Cr$_{0.18}$Ni$_{0.1}$ (AISI~304) based on electronic structure calculations performed within the density functional theory. In Sec.~\ref{theory} the computational methods used for the calculation of the electronic density of states (DOS) and for the electron heat capacity as well as the electron phonon coupling factor are described.  In Sec.~\ref{results} our results of the analysis of the thermophysical properties are compared with predictions for fcc Fe. In Sec.~\ref{summary} a brief summary of the results is given.
\section{Theory and methods \label{theory}}
\subsection{Electronic structure calculation} 
The electronic structure is obtained from ab initio calculations using the spin polarized relativistic Korringa-Kohn-Rostoker (SPR-KKR) band structure program package\cite{SPR-KKR6.3,korringa1}. In this method the Bloch waves and the corresponding electronic bands of a material are calculated by solving the Dirac equation using the Green function formalism. For our calculations a fully relativistic implementation of the KKR-formalism within framework of spin density functional theory is used\cite{korringa1,kohn1,ebert1}. To determine the electronic structure of Fe$_{0.72}Cr_{0.18}Ni_{0.1}$, e.g. stainless steel it is necessary to treat the system as a disordered type of a solid state crystal. Therefore the coherent potential approximation (CPA) was applied\cite{velicky1}. In terms of the CPA it is possible to represent a disordered or disturbed system as a hypothetical ordered effective medium. The resulting Green function of the disordered medium can be represented as configuration averaged Green function\cite{razee1}. In that sense a system of randomly distributed  disordered atoms is assumed. The calculation of the paramagnetic state above the Curie temperature has been performed in a disordered local moment (DLM) model\cite{gyorffy1}. It has been shown previously by Vitos~\textit{et~al.}\cite{Vitos2003} that for the correct determination of the electronic structure of stainless steel as well as the elastic properties the DLM model is appropriate. In this model the two spin directions of Fe are distributed on two potentials on one lattice site representing a disordered magnetic state. It is assumed that the effective host potential is occupied by 50\% electrons with spin up and 50\% electrons with spin down direction, hence the configuration averaged Green function of the effective medium includes the magnetic and the chemical disorder of Fe$_{0.72}$Cr$_{0.18}$Ni$_{0.1}$. 
The chemical disorder was considered due to the elements included according to their stoichiometric composition. For the calculation of the effective CPA medium the Mills algorithm was used. The exchange and correlation interaction of the electrons were described using the exchange correlation functional in the parametrization of Vosko, Wilk and Nusair\cite{vosko1}, i.e. the calculation were done applying the local spin density approximation (LSDA). 

For Fe$_{0.72}$Cr$_{0.18}$Ni$_{0.1}$ we used a face centered cubic (fcc) crystal structure with a lattice parameter of 3.59~\AA\cite{F.C.NascimentobC.M.Lepienskia} whereas for Fe in the fcc structure we used 3.52~\AA\cite{R2015220}. The convergence of the self-consistent potentials have been carefully checked with a number of 834~k-points in the irreducible part of the Brillouin zone.
 
\subsection{Electron heat capacity}
The electronic contribution to the specific heat considering a constant volume of a metal can be calculated by taking the partial differential over total electron energy density with respect to the electron temperature $C_e=\partial U/\partial T_e |_{V}$. The specific heat of the electron gas in dependence on the electronic temperature is described as\cite{Kittel,Ashcroft&Mermin}:
\begin{equation}
C_e(T_e) = \int \limits_{-\infty}^{+\infty} (E-E_F) \frac{\partial f(E,\mu,T_e)}{\partial T_e}g(E)\mathrm{d}E,
\label{eq:Ce}
\end{equation}
where $g(E)$ is the DOS at energy $E$, $f(E,\mu,T_e)$ is the Fermi$-$Dirac distribution function, $E_F$ is the Fermi energy and $\mu$ the chemical potential. The Fermi distribution, which gives the occupation number of a particular energy level, is defined as: 
\begin{equation} 
f(E,\mu,T_e)=\frac{1}{e^{(E-\mu)/k_B\cdot T_e}+1},
\end{equation} 
The evaluation of the electron heat capacity from Eq.~(\ref{eq:Ce}) requires the determination of the electronic DOS and the derivative of Fermi distribution function with respect to electron temperature. The temperature derivative of Fermi function is only non-zero near $E_F$. The determination of $\partial f/\partial T_e$ in Eq.~(\ref{eq:Ce}) at higher temperatures requires the evolution of chemical potential as a function of the electron temperature $\mu(T_e)$.

The chemical potential can be interpreted as characteristic energy, which defines the internal energy change of the system under conditions of constant entropy and volume, when one more particle is added \mbox{$\mu=\partial U/\partial N |_{S,V}$}\cite{R.Baierlein,G.CookR.H.Dickerson}. At a temperature of $0$~K all states below $E_F$ are filled whereas above $E_F$ all states are empty. Therefore the system must be in an energetic minimum. The entropy is related to the number of possible microstates and for a system containing a certain number of particles there exist only one with an energetic minimum. For this ground state the entropy is equal to zero\cite{Wilks}. After the addition of one particle the system must be in the ground state of new system for reaching of thermal equilibrium at absolute zero temperature. Thus, by adding one particle above $E_F$ the internal energy enhancement in system must be equal to $E_F$ and the entropy remain zero\cite{R.Baierlein}.

If the electron temperatures are essentially lower than the Fermi temperature, the chemical potential can be approximated by the Sommerfeld expansion for the free electron gas model\cite{Ashcroft&Mermin}. With the Sommerfeld expansion of the free electronic energy for metals the well-known linear temperature dependence of electron heat capacity can be derived. However, at high electron temperature the Sommerfeld theory of metals is not valid and the electron heat capacity can be calculated for exited electrons from Eq.~(\ref{eq:Ce}) by using a precise description of the electronic properties of the DOS.  The chemical potential $\mu(T_e)$ as a function of the electron temperature can be found directly by evaluating and iterating of integration from Eq.~(\ref{eq:Ne}) at various electron temperatures $T_e$. For the integration the value of the integral must be constant to the value of the electronic density $n_e = N_e/V$\cite{Ashcroft&Mermin}: 
\begin{equation}
n_e=\int\limits_{-\infty}^{+\infty}f(E,\mu(T_e),T_e)\cdot g(E)\mathrm {d}E.
\label{eq:Ne}  
\end{equation}

\subsection{Electron-phonon coupling factor}
The first theory of evaluating the electron-phonon coupling factor was done by Kaganov~\textit{et~al.}\cite{M.I.KaganovI.M.LifshitzandL.V.Tanatarov}. The model is based on the
electron-lattice energy exchange rate by consideration of the electron
relaxation time. Under the condition that the electron temperature is equal to
the lattice temperature, a constant coupling factor can be obtained. In
the next step a model was introduced by Chen~\textit{et~al.}, in which
electron-electron and electron-phonon relaxation times are included\cite{J.K.ChenW.P.LathamJ.E.Beraun}. The constant value of electron-phonon
coupling factor estimated by Kaganov and the linear temperature dependence proposed by
Chen are limited to low electron temperature and are not valid for high electron
temperature. At high electron temperatures, however, the thermal excitation
must be considered. The electrons located below the $E_F$ begin to
contribute to the electron-phonon energy exchange rate. The free electron gas
model cannot be applied and  the full spectrum of
the electronic DOS for metals is required.

The model for the calculation of the electron-phonon coupling factor is based on a general
description of the electron-phonon energy exchange rate with two distinct
temperatures developed by Allen\cite{P.B.Allen}. Wang~\textit{et~al.} extended
this theoretical model by inclusion of $d$-band states for thermal excitation at
higher electron temperatures in order to get accurate results for gold 
in comparison with time-resolved electron temperature measurements\cite{W.Y.WangD.MarkRiffeY.S.LeeandM.C.Downer}. In his work Lin~\textit{et~al.} has
rewritten the theory corresponding to the electron-phonon coupling factor by merging
the theory of Allen and Wang to compare it with the free electron gas model of
various transition metals\cite{Lin&Zhigilei&Celli}. The temperature dependence
of electron-phonon coupling factor within this approach can be expressed as
\begin{equation}
G(T_e) = \frac{\pi \hslash k_B \lambda \langle \omega^2
\rangle}{g(E_F)} \int \limits_{-\infty}^{+\infty} g^2(E) \left(-\frac{\partial
f}{\partial E}\right) \mathrm{d}E,
\label{eq:G}
\end{equation}
where $\lambda$ is the dimensionless electron-phonon mass enhancement parameter \cite{G.Grimval}
and $\langle\omega^2\rangle$ is the second moment of the phonon spectrum
defined by McMillan \cite{W.L.McMillan}, and $g(E_F)$ is the electron DOS at
the Fermi level $E_F$.  At low electron temperature $-\partial f/\partial E$
reduces to a delta function centered at $E_F$ and Eq.~(\ref{eq:G}) yields
a constant value for the electron-phonon coupling factor. Whereas at high electron
temperatures the delta function is shifting away from
$E_F$. This significant shift of $E_F$ causes a temperature dependence of the electron-phonon coupling factor.
The material parameter  $\lambda\langle\omega^2\rangle$ can be obtained
by using the calculated or experimental value of $\lambda$\cite{S.Y.SavrasovD.Y.Savrasov} and the approximation $\langle\omega^2\rangle=\theta^2 _D /2$\cite{D.A.Papaconstantopoulos}. 

Sacchetti has proposed a simple model for treating the electron-phonon
interaction in binary alloys by applying the
CPA to calculate the specific heat enhancement. Within this approximation in
random binary alloys a description for electron-phonon enhancement factor is
developed. It is given by the average of the local enhancement factors considering
a weak-scattering system $\lambda=x\lambda_A+y\lambda_B$\cite{F.Sacchetti}.

In absence of $\lambda$ for ternary Fe alloys a quantitative estimation can be developed by extending Saccetti's approximate description to threee components
\begin{equation}
\lambda = x\lambda_A + y\lambda_B + z\lambda_C.
\label{eq:lambda} 
\end{equation}
\section{Results \label{results}}
\subsection{Ground state properties of Fe and Fe$_{0.72}$Cr$_{0.18}$Ni$_{0.1}$}
\begin{figure}[htb]
\includegraphics[width=0.55\textwidth]{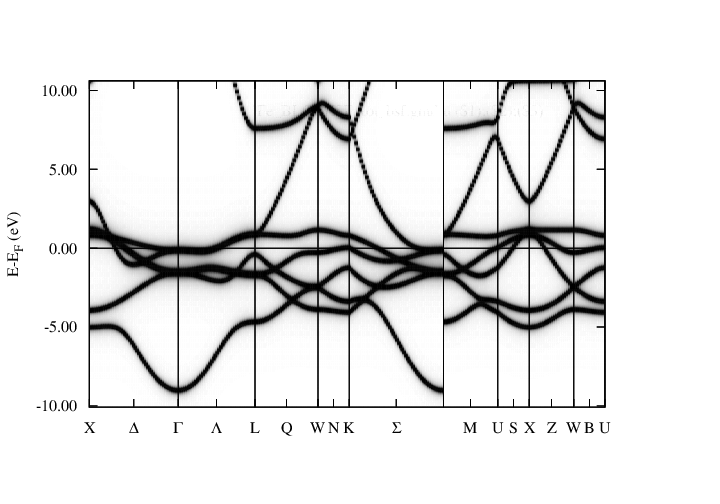}
\caption{\label{fig:Fe_fcc_bandstructure} Bloch spectral function of fcc Fe calculated using the SPR-KKR program package \cite{SPR-KKR6.3}. The energy is denoted with respect to $E_F$.} 
\end{figure}

\begin{figure}[htb] 
\includegraphics[width=0.55\textwidth]{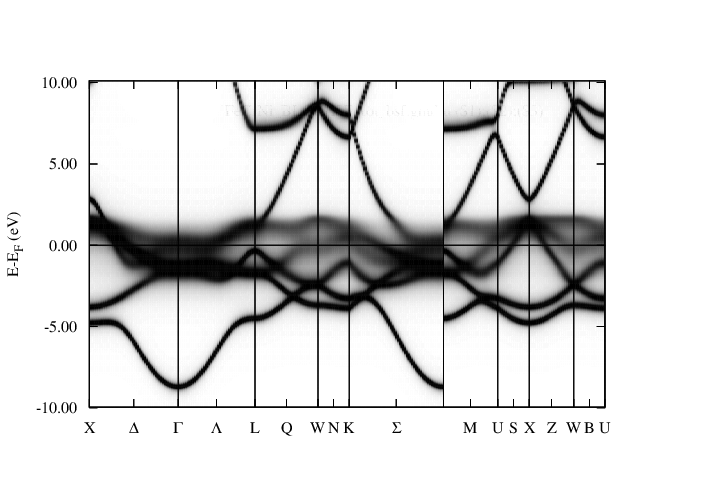}
\caption{\label{fig:FeCrNi_fcc_bandstructure} Bloch spectral function of fcc Fe$_{0.72}$Cr$_{0.18}$Ni$_{0.1}$ calculated using the SPR-KKR
program package \cite{SPR-KKR6.3}. The energy is denoted with respect to $E_F$.} 
\end{figure}

The ground state properties of Fe and Fe$_{0.72}$Cr$_{0.18}$Ni$_{0.1}$
are the basis of the additional calculations for the thermal properties.
In Figs.~\ref{fig:Fe_fcc_bandstructure} and \ref{fig:FeCrNi_fcc_bandstructure}
the Bloch spectral function of bulk fcc Fe and Fe$_{0.72}$Cr$_{0.18}$Ni$_{0.1}$ are shown.
The Bloch spectral function can be interpreted as k-resolved DOS in reciprocal space.
There are mainly two types of bands visible. For the energy range from
-10 eV to -5 eV a $s$-band dominates the k-path from X to $\Gamma$. Around
$E_F$ detailed bands occur coming from the $d$-states of Fe. 
These bands essentially define the electronic and magnetic properties
of Fe and Fe$_{0.72}$Cr$_{0.18}$Ni$_{0.1}$. 
For Fe$_{0.72}$Cr$_{0.18}$Ni$_{0.1}$ a broadening of the $d$-bands applies,
whereas the $s$-band is unaffected. The broadening of electronic bands
due to chemical and magnetic disorder is a well known effect that
has been investigated in previous works\cite{LP06,M04}. This has
an impact on the magnetic and electronic properties as well as
the thermal properties important for the laser ablation.
\subsection{Thermal properties of Fe and Fe$_{0.72}$Cr$_{0.18}$Ni$_{0.1}$} 
\begin{figure*}[htb]%
    \subfigure[\label{fig:DOS}]{
    \includegraphics[width=0.47\textwidth]{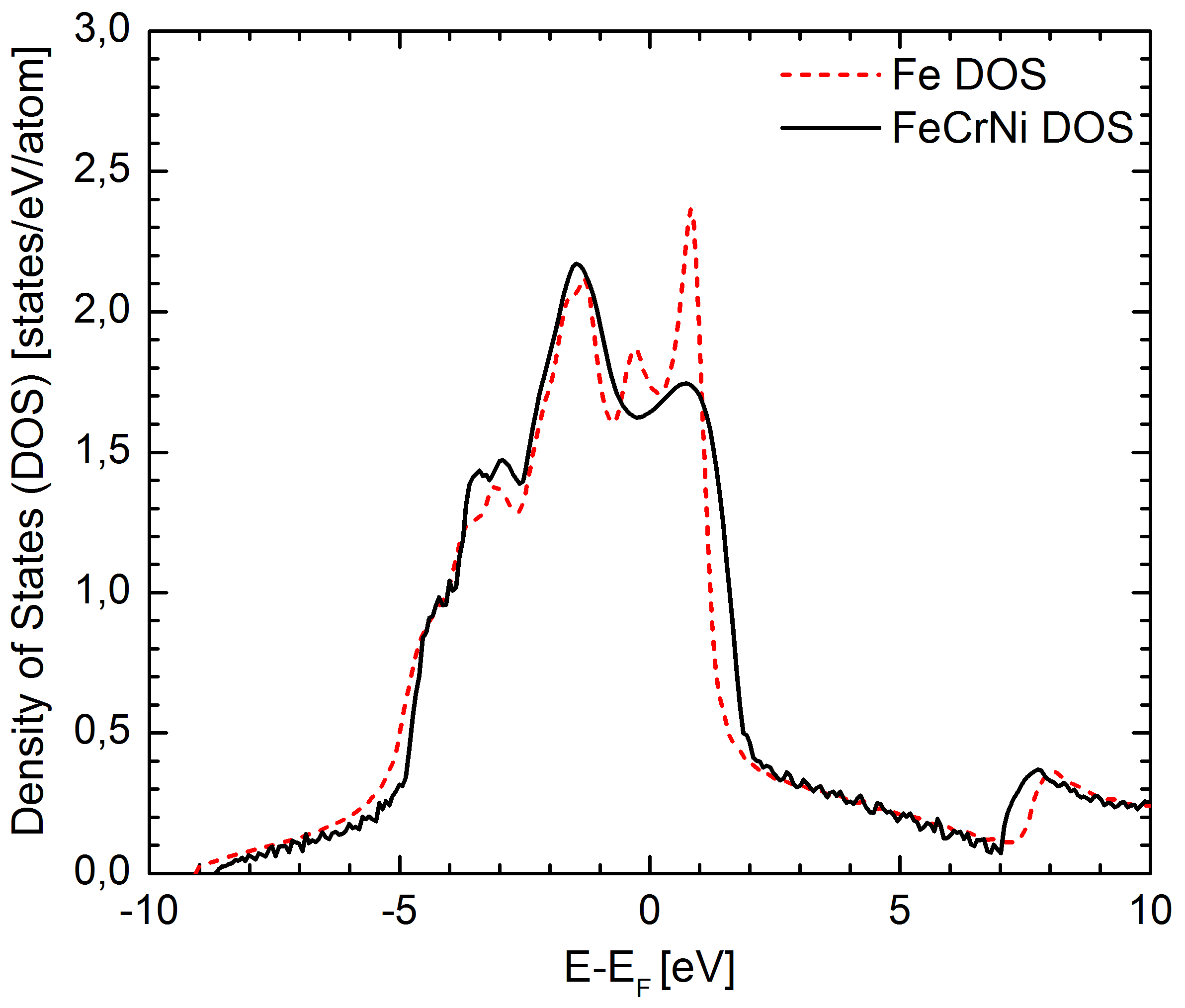}
    }%
    \quad
    \subfigure[\label{fig:mu}]{
    \includegraphics[width=0.47\textwidth]{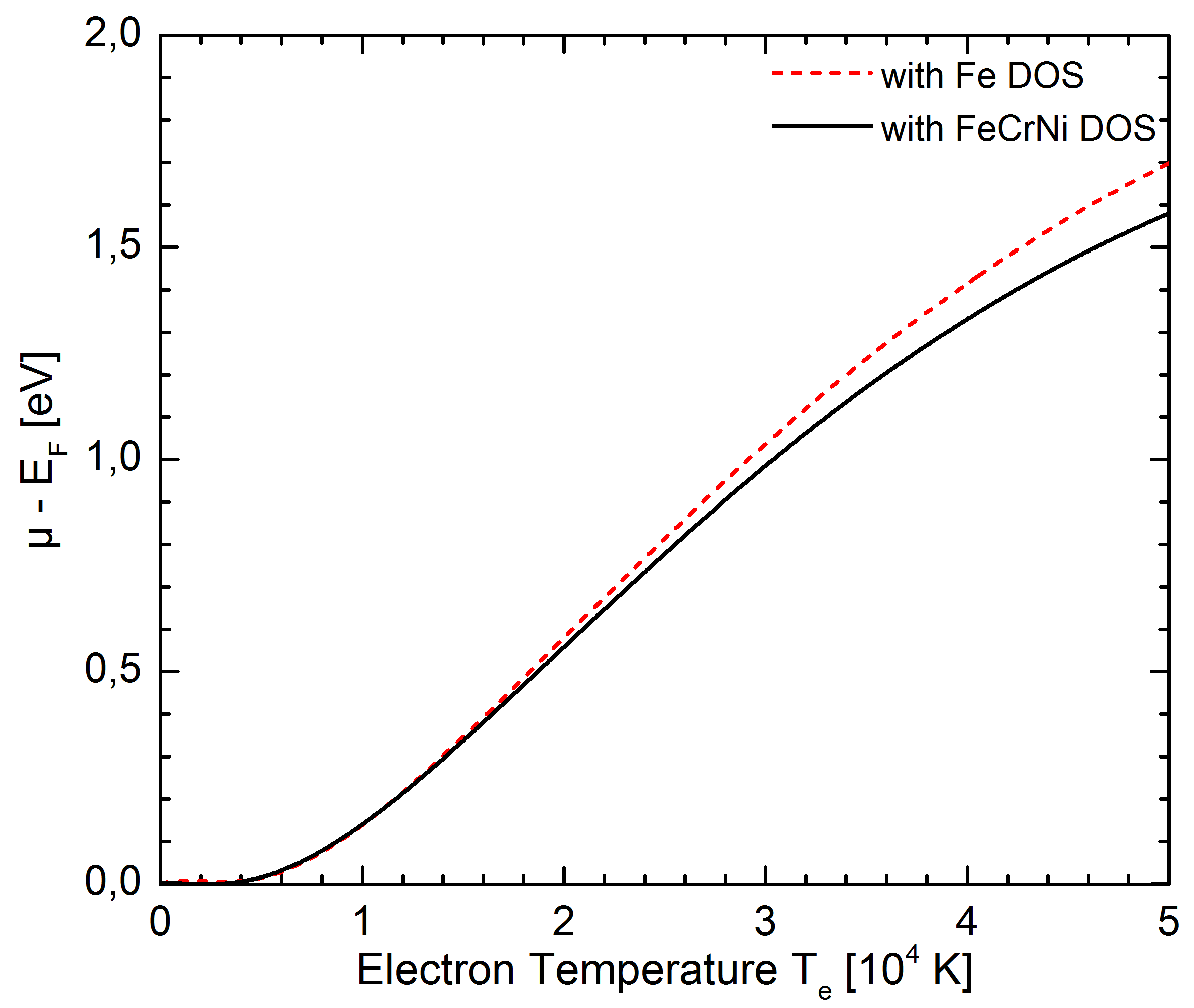}
    }%
    \\
    \subfigure[\label{fig:Ce}]{
    \includegraphics[width=0.47\textwidth]{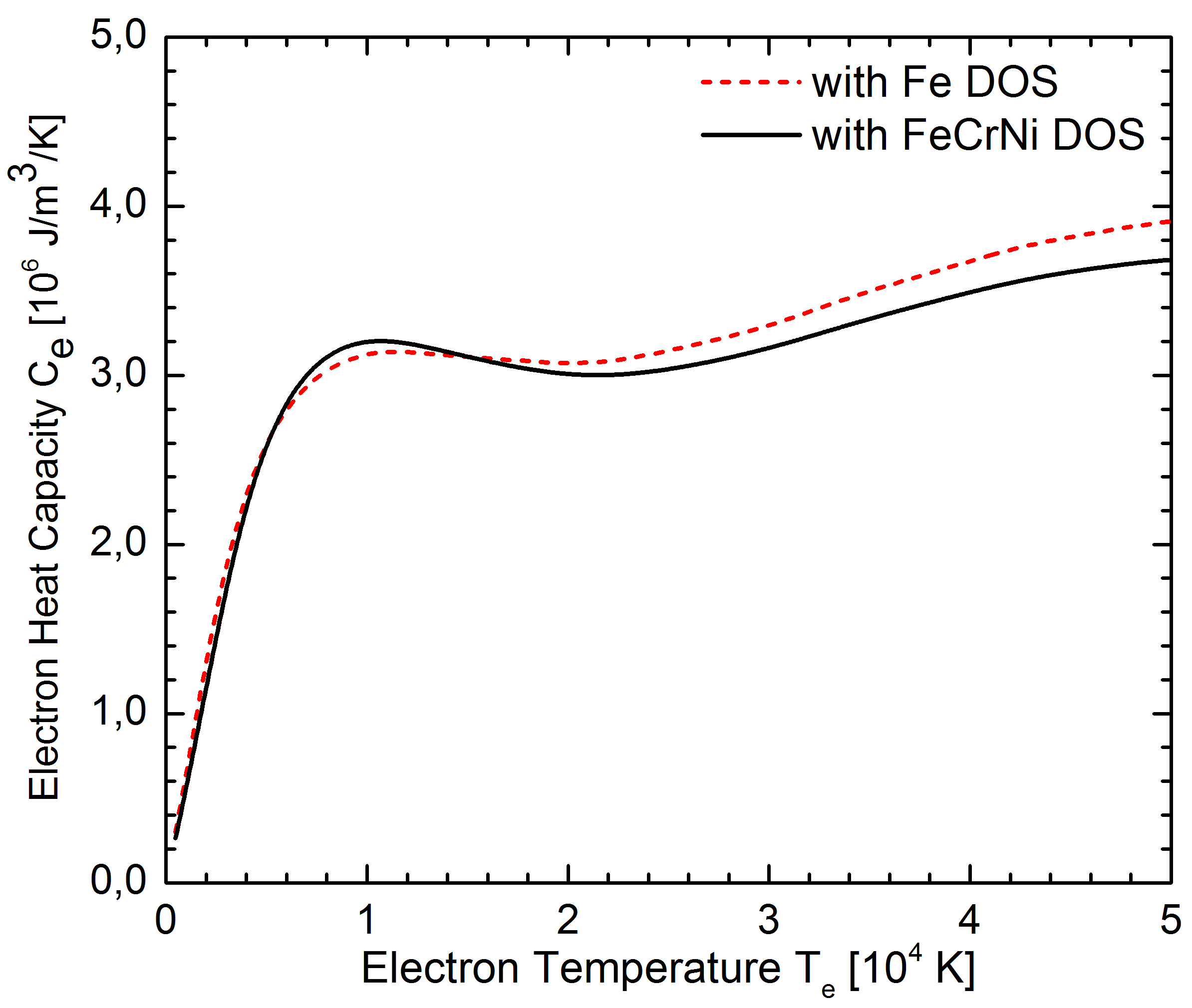}
    }%
    \quad
    \subfigure[\label{fig:G}]{
    \includegraphics[width=0.47\textwidth]{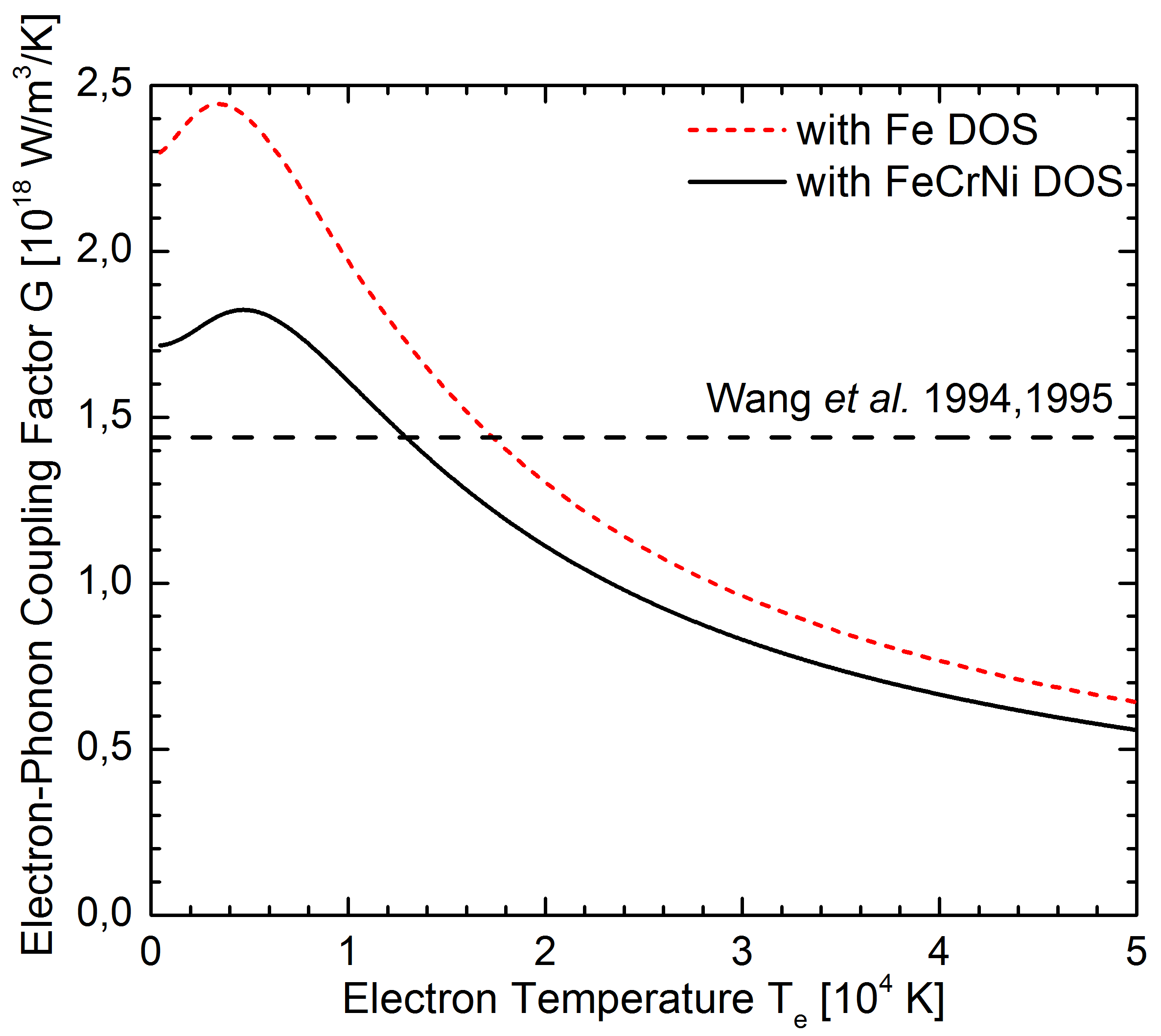}
    }%
\caption{(a) The electron DOS calculation for Fe$_{0.72}$Cr$_{0.18}$Ni$_{0.1}$
obtained at a temperature of 0 K from SPR-KKR electronic sturcture package. The energy is denoted with respect to the Fermi energy, $E_F$, at 0~K, (b) the chemical potential
$\mu$ as a function of the electron temperatures. (c) The electron heat capacity
$C_e$, (d) the electron-phonon coupling factor $G$ as a function of the electron
temperature using the material parameter $\lambda \langle \omega^2 \rangle = 217.75$
$meV^2$ for Fe and $\lambda \langle \omega^2 \rangle = 184.44$ $meV^2$ for
Fe$_{0.72}$Cr$_{0.18}$Ni$_{0.1}$} 
\end{figure*}

In the following the calculations for the DOS as well as for the thermal properties of Fe$_{0.72}$Cr$_{0.18}$Ni$_{0.1}$ are presented. The electronic structure calculations of the ground state are performed with the SPR-KKR package for a realistic description of the paramagnetic properties. The electron heat capacity and the electron-phonon coupling factor are compared between Fe and Fe$_{0.72}$Cr$_{0.18}$Ni$_{0.1}$. In Fig. \ref{fig:DOS} the electronic DOS for Fe$_{0.72}$Cr$_{0.18}$Ni$_{0.1}$ (fcc) and Fe (fcc) indicate similar characteristics of the electronic DOS. In the range of -9 eV and -5 eV for Fe$_{0.72}$Cr$_{0.18}$Ni$_{0.1}$ and Fe a $s$-band can be identified. A major common feature is the presence of a high DOS in the range between -5 eV and 2 eV. This region can be associated with the occurrence of a $d$-band. In the case of Fe$_{0.72}$Cr$_{0.18}$Ni$_{0.1}$ the $d$-band reaches $\sim 2$~eV above Fermi energy level and thus is not occupied. However, the $d$-band edge of Fe is located $\sim 1$~eV closer to $E_F$. At energies $\sim 1$ eV and $\sim 2$ eV  below $E_F$ a distinct characteristic peak of the electronic DOS of Fe$_{0.72}$Cr$_{0.18}$Ni$_{0.1}$ vanishes in comparison to Fe. As a result the $d$-band of Fe has a higher occupation number compared to Fe$_{0.72}$Cr$_{0.18}$Ni$_{0.1}$. With that a larger number of electrons can be excited into the conduction band resulting in a higher influence on the electronic heat capacity $C_e$ and the electron-phonon coupling factor $G$.

The predicted temperature dependence of the chemical potential for Fe and
Fe$_{0.72}$Cr$_{0.18}$Ni$_{0.1}$ is shown in Fig.~\ref{fig:mu}. At low electron temperature
$\sim 10^3$~K with respect to energy $\sim 0.1$ $eV$ the thermal excitation of
the $s$-band is the dominant effect. The chemical potential remains nearly
constant because the internal energy change is equal to $E_F$ and the
entropy remains zero. At higher electron temperatures, however,
the electrons at the energy levels below the $E_F$ can be easily exited. The electronic excitation 
to the conduction band leads to a shift of the chemical
potential to higher energies [see Fig.~\ref{fig:mu}]. 
The excitation of electrons change the number of
occupied states below and above $E_F$. This alteration of occupied states
leads to a change of the internal energy. However the entropy of the system must
not increase. As a result the chemical potential must shift to higher energies
so that the internal energy does not increase and hence the entropy
remains zero. The shift of the chemical potential is a consequence of the change
in the electronic occupation by thermal excitation. Because of that the fundamental 
distinction of the chemical potential for Fe$_{0.72}$Cr$_{0.18}$Ni$_{0.1}$ and Fe can be
reduced to different $d$-band widths and as a consequence the number
of accessible excited states. 

Fig.~\ref{fig:Ce} shows the temperature dependence of the electron heat
capacity. The electron heat capacity of Fe and of Fe$_{0.72}$Cr$_{0.18}$Ni$_{0.1}$ 
is calculated by inserting the results for
the chemical potential and the DOS of Fe and Fe$_{0.72}$Cr$_{0.18}$Ni$_{0.1}$ into Eq.~(\ref{eq:Ce}).
For electron temperatures $\sim 5 \cdot 10^4$~K the electron heat capacity
of Fe and Fe$_{0.72}$Cr$_{0.18}$Ni$_{0.1}$ follows almost a linear behavior up to
a local peak at $\sim 1 \cdot 10^4$~K. The
predicted linear dependence is affected by excitations from
the occupied states of the $d$-band close to $E_F$.
The number of
excited electrons are determined by the width of thermal excitation ($\sim k_B\cdot T_e$) in an energy interval around $E_F$ and the increase of the number
of conductive electrons in the $s$-band. With an increase of the electron temperature above
$\sim 5 \cdot 10^3$~K the chemical potential shift to higher energies. This creates
a slightly drop of the curve till a minimum with a subsequent increase of heat capacity
above electron temperature $\sim 2.4 \cdot 10^4$~K. At higher electron
temperatures a deviation of the electron heat capacity about $5~\%$ comparing Fe and
Fe$_{0.72}$Cr$_{0.18}$Ni$_{0.1}$ occurs [see Fig.~\ref{fig:Ce}]. This results from
the shift of the chemical potential to the $d$-band edge. The number of possible
states of Fe located at $\sim 1$~eV and $\sim 0.5$~eV
above $E_F$ can contribute directly to the electron heat capacity by thermal
excitation. This exhibits a decline in the electron heat capacity
of Fe$_{0.72}$Cr$_{0.18}$Ni$_{0.1}$ at high electron temperatures. With that
a low deviance of stored electron heat with respect to the change of the
electron temperature
is revealed between Fe and Fe$_{0.72}$Cr$_{0.18}$Ni$_{0.1}$. Thus, the predicted trend of
the electron heat capacity $C_e$ causes a similar transient evolution of the
electron temperature in the electron subsystem of Fe and Fe$_{0.72}$Cr$_{0.18}$Ni$_{0.1}$ during the time
of the electron-lattice non-equilibrium processes induced in a metal target by
the fast laser energy deposition with ultra-short laser pulses.

The electron heat capacity is also sensitive to the electronic structure
calculation of the DOS discussed above in either instance for Fe$_{0.72}$Cr$_{0.18}$Ni$_{0.1}$ and Fe. The
excitation of $d$-band electrons has a strong effect for the temperature
dependence of the electron heat capacity and must be taken into account in
the estimation of the electron-phonon coupling. The calculation of the electron
temperature dependence of the electron-phonon coupling is performed with
Eq.~(\ref{eq:G}), which requires the information of $\lambda
\langle \omega^2 \rangle$. The value for Fe$_{0.72}$Cr$_{0.18}$Ni$_{0.1}$, 
$\lambda \langle \omega^2\rangle = 184.44$~meV$^2$, is estimated by approximation from
Eq.~(\ref{eq:lambda}) and $\langle \omega^2 \rangle = \theta^2 _D /2$ using
experimental measurements of the Debye temperature $\theta _D$ for AISI 304 at T
$=0$ K \cite{Ledbetter&Weston&Naimon}. In Table~\ref{table:a<w>} the values of $\lambda$ calculated by self-consistent band structure methods and the measured Debye
temperatures for Fe, Cr, Ni are listed. 

\begin{table}[ht] \caption{The material parameter  $\lambda \langle \omega^2\rangle$
estimated by using the calculated value of $\lambda$
\cite{D.A.Papaconstantopoulos} and the approximation $\langle \omega^2
\rangle = \theta^2 _D /2$ meV$^2$. $\lambda$ is the dimensionless
electron-phonon mass enhancement parameter and  $\theta_D$  is the Debye
temperature $(K)$ taken from Papaconstantopoulos~\textit{et~al.}\cite{D.A.Papaconstantopoulos}.}
\begin{ruledtabular}
\begin{tabular}{ccccc}
& Fe & Cr & N & FeCrNi \\ \hline \\
$\lambda$ & 0.270 & 0.131 & 0.084 &
0.226 \\ $\theta_D$ & 467 & 630 & 450 & 468 \\
$\lambda \langle \omega^2 \rangle$ & 217.75 & 192.27 & 62.90 & 183.44 \\ [1ex]
\end{tabular} \label{table:a<w>}
\end{ruledtabular}
\end{table}
The electron temperature dependence of the electron-phonon coupling factor $G(T_e)$
predicted for Fe and Fe$_{0.72}$Cr$_{0.18}$Ni$_{0.1}$ is very similar and exhibits the same
qualitative features as shown in Fig.~\ref{fig:G}. The
features of both dependencies are an increase of the electron-phonon
coupling factor up to $\sim 4 \cdot 10^3$~K, followed by a decrease with further
temperature rise. The rise of the electron-phonon
coupling factor at electron temperatures below $\sim 4 \cdot 10^3$~K can be explained
by thermal excitations of a large number of $d$-band electrons, which contribute
to the electron-phonon collision and accordingly to energy exchange. The increase
of the electron-phonon coupling factor indicates a faster energy transfer by
the electron-phonon scattering process from hot electrons to the lattice within the
framework of TTM. 

The start value of $G$ for Fe$_{0.72}$Cr$_{0.18}$Ni$_{0.1}$ at low electron temperatures is reduced
to $ \sim 1.7 \cdot 10^{18}$~W/m$^3$/K. A possible reason can be a lower value
of  $\lambda$ at similar values of $\langle \omega^2 \rangle$. This decrease of
the electron-phonon interaction leads to a lower collision frequency and
therefore to smaller energy exchange between electron and lattice at lower
electron temperatures. At higher temperatures above $5 \cdot 10^3$~K the
excitation of $d$-band electrons exceeds the high density of the $d$-band edge of
Fe located at $1$~eV [see Fig.~\ref{fig:DOS}]. This shift reduces the
contribution of the $d$-band electrons to the electron-phonon coupling and
leads to a rapid drop of the energy exchange rate for Fe. In contrast for Fe$_{0.72}$Cr$_{0.18}$Ni$_{0.1}$
the $d$-band edge is located at an energy of $\sim 2$~eV. This leads to a smoother
decrease of the electron phonon coupling. With a temperature increase up to 
$\sim 5\cdot 10^4$~K a slower decrease of the electron-phonon coupling can be seen.
The predicted temperature dependencies can be interpreted as equilibrium of the
excited electrons coming from deeper energies of the $d$-band
and the continuing the shift of the chemical potential to higher energies towards
$d$-band edge. 
The calculated value of the electron-phonon
coupling factor for Fe at room temperature, shown in  Fig.~\ref{fig:G} ($2.3 \cdot 10^{18}$~W/m$^3$/K)
is higher than the value obtained
from previous swift-heavy-ion irradiation experiment ($1.44 \cdot 10^{18}$~W/m$^3$/K)
\cite{Z.G.WangCh.DufourE.PaumiertM.Toulemonde}.

The constant value of electron-phonon coupling strength ($1.44~\cdot~10^{18}$~W/m$^3$/K) has been estimated from comparison of molten phase radius around the
ion path in dependence on electronic energy loss between calculated and
experimental latent track radii irradiated by swift heavy ions within
the thermal-spike model. This value is smaller by a factor of $1.6$ to our prediction.
Wang $et \ al.$ assumed a uncertainty of $30 \%$ resulting from the input parameters
and the approximation of the electronic heat capacity and thermal conductivity
within the free electron gas theory for the calculation of the molten track radii.
The determination of the electron-phonon coupling constant can only be
considered as a rough estimation due to these discrepancies.
\cite{Z.G.WangCh.DufourE.PaumiertM.Toulemonde}.

The electron-phonon coupling factor is calculated for a wide
electron temperature range assuming a strong electron-phonon
non-equilibrium. In contrast in \cite{Z.G.WangCh.DufourE.PaumiertM.Toulemonde} 
the reported value of the electron-phonon coupling factor is estimated
by the calculation of molten
track radii. 
In this respect, the value of the electron-phonon coupling factor ($1.44~\cdot~10^{18}$~W/m$^3$/K) can be considered as an "effective" electron-phonon coupling constant and therefore be in a reasonable agreement with our calculations for Fe.

For the analysis of a laser energy deposition in a
metal alloy a temporal change of the thermophysical properties of
Fe$_{0.72}$Cr$_{0.18}$Ni$_{0.1}$ in comparison to Fe during the time of electron-phonon
equilibration can appear. Considering Fe$_{0.72}$Cr$_{0.18}$Ni$_{0.1}$ a decrease of the electron-phonon coupling
strength in a TTM simulation leads to a reduced energy transfer
from hot electrons to the lattice.
Consequently, a lower
transient evolution of the lattice temperature after irradiating with ultra-short
laser pulses could lead to an increase of the threshold
influence for the surface melting or laser ablation for Fe$_{0.72}$Cr$_{0.18}$Ni$_{0.1}$.

\section{Summary \label{summary}}
In this article the first calculation of the electron heat capacity and the electron-phonon coupling factor dependence on the electron temperature are presented for the stainless steel alloy Fe$_{0.72}$Cr$_{0.18}$Ni$_{0.1}$ (AISI 304). The calculation of electronic thermophysical parameters are based on detailed analysis of the electronic DOS obtained from the Munich SPR-KKR band structure program. With that we determined the properties of the electronic distribution for a randomly disordered system modeling Fe$_{0.72}$Cr$_{0.18}$Ni$_{0.1}$. The electron heat capacity and the electron-phonon coupling factor of Fe$_{0.72}$Cr$_{0.18}$Ni$_{0.1}$ show an increase affection by the thermal excitation in a wide range of the $d$-band with a high DOS at electron temperatures below 5$ \cdot 10^3$ K. By exceeding $10^4$~K a non-equilibrium of the excited $d$-band electrons and a reoccupation induced by shifting of the chemical potential to higher energies leads to a strong decrease of the electron-phonon coupling, whereas the electron heat capacity remains at considerably constant high values. The comparison with Fe indicates a similar qualitative and quantitative trend of the electron heat capacity of Fe$_{0.72}$Cr$_{0.18}$Ni$_{0.1}$ attributable to the high Fe content of 72 \% in the alloy. The negative deviation of electron-phonon coupling between Fe$_{0.72}$Cr$_{0.18}$Ni$_{0.1}$ and Fe by $\sim 25 \%$ is  due to electron-phonon interaction. Overall, a good agreement between the calculation of the electron heat capacity and the electron-phonon coupling factor of Fe$_{0.72}$Cr$_{0.18}$Ni$_{0.1}$ and Fe are observed.\newpage

\section{Acknowledgments}

Financial support of this work is provided by the funding from the European Union's Seventh Framework Programme (FP7/2007-2013) under grant agreement No. $310220$, from FORWIN program of MUAS,
CENTEM (CZ.1.05/2.1.00/03.0088), CENTEM PLUS (L01402) and the Munich School of Engineering within the framework of the TUM Applied Technology Forum. The authors would like to thank Dr. Gerhard Heise at Munich University of Applied Sciences for his valuable comments.

\bibliographystyle{apsrev4-1}
\bibliography{references}
\end{document}